\documentclass[runningheads]{svmult}

\usepackage{makeidx}   
\usepackage{graphicx}  
\usepackage{subeqnar}  
\usepackage{multicol}  
\usepackage{physprbb}  
\makeindex             

\newcommand{\greeksym}[1]{{\usefont{U}{psy}{m}{n}#1}}

\newcommand{\uDelta}{\mbox{\greeksym{D}}}

\newcommand{\fig}[1]{Fig.~\ref{#1}}

\newcommand{\eq}[1]{Eq.~(\ref{#1})}

\newcommand{\brefs}[1]{Refs.~\cite{#1}}

\renewcommand{\d}{{\rm d}}

\newcommand{\ub}{{\rm ub}}
\newcommand{\bd}{{\rm b}}
\newcommand{\piad}{\pi_{\rm ad}}

\begin{document}
\title*{Traffic of Molecular Motors}
\toctitle{Traffic of Molecular Motors}
%
%
\titlerunning{Traffic of Molecular Motors}
%
\author{Stefan Klumpp, Melanie J. I. M\"uller \and Reinhard Lipowsky }
\authorrunning{S. Klumpp, M. M\"uller and R. Lipowsky}
%
%
\institute{Max-Planck-Institut f\"ur Kolloid- und Grenzfl\"achenforschung, \\
  Wissenschaftspark Golm, 14424 Potsdam, Germany}

\maketitle              

\begin{abstract}
  Molecular motors perform active movements along cytoskeletal
  filaments and drive the traffic of organelles and other cargo
  particles in cells. In contrast to the macroscopic traffic of cars,
  however, the traffic of molecular motors is characterized by a
  finite walking distance (or run length) 
  after which a motor unbinds from the
  filament along which it moves. Unbound motors perform Brownian
  motion in the surrounding aqueous solution until they rebind to a
  filament.  We use variants of driven lattice gas models to describe
  the interplay of their active movements, the unbound diffusion, and
  the binding/unbinding dynamics. If the motor concentration is large,
  motor-motor interactions become important and lead to a variety of
  cooperative traffic phenomena such as traffic jams on the filaments,
  boundary-induced phase transitions, and spontaneous symmetry
  breaking in systems with two species of motors. If the filament is 
  surrounded by a large reservoir of motors, the jam length, i.e., the 
  extension of the traffic jams is of the order of the walking distance. 
  Much longer jams can be found in confined geometries such as 
  tube-like compartments.
\end{abstract}

\section{Introduction}

The traffic of vesicles, organelles, protein complexes, messenger RNA,
and even viruses within the cells of living beings is driven by the
molecular motors of the cytoskeleton which move along cytoskeletal
filaments in a directed fashion
\cite{Howard2001,Schliwa2003,Schliwa_Woehlke2003}. There are three
large classes of cytoskeletal motors, kinesins and dyneins which move
along microtubules, and myosins which move along actin filaments.
These motors use the free energy released from the hydrolysis of
adenosinetriphosphate (ATP), which represents their chemical fuel, for
active movement and to perform mechanical work. They move in discrete
steps in such a way that one molecule of ATP is used per step.
Typical step sizes are $\sim 10$~nm, typical motor velocities are in
the range of $\mu$m/sec.

Since the interior of cells is quite crowded and motors are strongly
attracted by the filaments, which leads to relatively large motor 
densities along the filaments, it is interesting to study the 
collective traffic
phenomena which arise from motor--motor interactions, in particular
the formation of traffic jams due to the mutual exclusion of motors
from filament sites. To study these cooperative phenomena
theoretically we have introduced new variants of driven
lattice gas models \cite{Lipowsky__Nieuwenhuizen2001} which have been
studied extensively during the last years both by our group
\cite{Lipowsky__Nieuwenhuizen2001,Nieuwenhuizen__Lipowsky2002,Klumpp_Lipowsky2003,Klumpp_Lipowsky2004,Nieuwenhuizen__Lipowsky2004,Klumpp__Lipowsky2005,Lipowsky_Klumpp2005,Mueller__Lipowsky2005}
and by several other groups
\cite{Parmeggiani__Frey2003,Evans__Santen2003,Popkov__Schuetz2003,Klein__Juelicher2005,Arizmendi__Family2005,Nishinari__Chowdhury2005}
and which will be described below. These models are related to lattice
gas models for driven diffusive systems and exclusion processes as 
studied in the context of non-equilibrium phase
transitions
\cite{Katz__Spohn1984,Krug1991,Kolomeisky__Straley1998,Kafri__Toeroek2002}
and highway traffic
\cite{Nagel_Schreckenberg1992,Chowdhury__Schadschneider2000}. Since
molecular motors can be studied in a systematic way using biomimetic
systems which consist of a small number of components (such as motors,
filaments, and ATP), they can also serve as model systems for the
experimental investigation of driven diffusive systems.

Although the traffic of cargo particles pulled by molecular motors
within cells is remarkably similar to the macroscopic traffic on streets
or rails, there is an important difference which is a direct consequence of
the nanoscale size of molecular motors: The motor--filament binding
energy can be overcome by thermal fluctuations which are ubiquitous on
this scale, and molecular motors therefore have a finite walking
distance or run length after which they unbind from the filament 
along which they
move. This walking distance is typically of the order of 1~$\mu$m for
a single motor molecule.\footnote{In order to transport a cargo
  actively over larger distances as, e.g., in the axon of a nerve
  cell, several motors work together in a cooperative fashion. We have
  recently shown that 7--8 motors are sufficient for processive
  transport over distances in the centimeter range as necessary in
  axons \cite{Klumpp_LipowskyCoopTr}.}  Likewise, unbound motors which
diffuse freely in the surrounding aqueous solution, can bind to a
filament and start active movement. In contrast to highway traffic,
where additional cars enter only at on-ramps, i.e.\ at specific
locations, binding of molecular motors occurs along the full length of
the filaments.  In addition to stepping along a one-dimensional track
and mutual exclusion, lattice models for the traffic of molecular
motors must therefore also describe the dynamics of motor--filament
binding and unbinding as well as the diffusive movement of the unbound
motors.\footnote{These processes have not been taken into account in
  earlier studies of exclusion effects in many-motor systems which
  were based on ratchet models
  \cite{Derenyi_Vicsek1995,Derenyi_Ajdari1996,Aghababaie__Plischke1999}.}

In contrast to the transport properties of single motor molecules
which have been studied extensively during the last 15 years
\cite{Howard2001,Schliwa2003}, the traffic phenomena in many-motor
systems have only recently attracted the interest of experimentalists
and are still largely unexplored from the experimental point of view.
The quantity of main interest has so far been the profile of the bound
motor density along a filament. Density profiles with a traffic
jam-like accumulation of motors at the end of filaments have
been observed in vivo for a kinesin-like motor which was overexpressed in
fungal hyphae \cite{Konzack2004,Konzack__Fischer2005}. Recently, motor
traffic jams have also been observed in biomimetic in vitro systems
using both conventional kinesin (kinesin 1) \cite{Leduc__Prost2004}
and the monomeric kinesin KIF1A (kinesin 3)
\cite{Nishinari__Chowdhury2005}.

In the following, we will give a short overview over the lattice
models for molecular motors and discuss the motor traffic in various
systems which differ mainly in the compartment geometry and the
arrangement of filaments. 
In section \ref{jams}, we address the length of motor jams on filaments 
and argue that, in the presence of a large motor reservoir this jam 
length is typically of the order of the walking distance. Longer jams 
are found in confined geometries as discussed in section \ref{closed}. 
In the last section of the paper, we briefly review our results for
systems with two motor species.

\section{Lattice models for molecular motor traffic}

To describe the interplay of the movements of bound and unbound
motors, we have introduced a class of lattice models which incorporate
the active movement of bound motors, the passive diffusion of unbound
motors, and the motor--filament binding and unbinding dynamics
\cite{Lipowsky__Nieuwenhuizen2001}. These models can also account for
motor--motor interactions such as their mutual exclusion from binding
sites of the filament.

\begin{figure}[tb]
  \centering
  \includegraphics[angle=0,width=0.8\columnwidth]{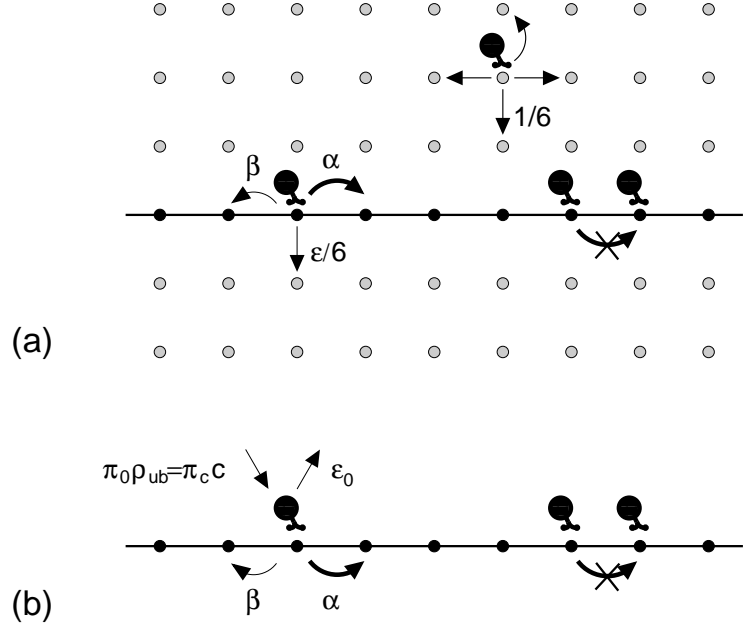}
  \caption{Lattice model for molecular motor traffic: (a) Molecular motors
    step in a biased fashion along a filament (black line).  With
    probability $\epsilon/6$, a motor
    unbinds from the filament by stepping to an adjacent non-filament
    site. Unbound motors perform symmetric random walks and, when
    reaching a filament site, rebind to it with probability $\piad$.
    Mutual exclusion implies that motors cannot step to lattice sites
    that are already occupied by another motor. (b) In some situations, the
    unbound motor density $\rho_\ub$ (or the corresponding
    concentration $c$) can be considered as constant. In that case,
    bound motors move along the filament as in (a) and unbind from it
    with probability $\epsilon_0=n_{\rm ad}\epsilon/6$. Binding of a motor to
    an empty filament site occurs with probability 
    $\pi_0\rho_\ub=n_{\rm ad}\piad\rho_\ub/6$
    which can also be expressed as $\pi_c c$ as discussed in section
    \ref{jams}.}
  \label{fig:latticemodel}
\end{figure}

We describe the motor movements as random walks on a (in general,
three-dimensional) cubic lattice as shown in
\fig{fig:latticemodel}(a).  Certain lines on this lattice represent
the filaments. The lattice constant is taken to be the motor step size
$\ell$ which for many motors is equal to the filament periodicity. 
When a motor is localized at a filament site, it performs a biased 
random walk. Per unit time
$\tau$, it makes forward and backward steps with probabilities
$\alpha$ and $\beta$, respectively. With probability $\gamma$, the
motor makes no step and remains at the same site. The latter parameter
is needed to account for the fact that if the lattice constant is
given by the motor step size, unbound diffusion over this scale is
much faster than an active step of a bound motor. Finally, the motor
hops to each of the adjacent non-filament sites with probability
$\epsilon/6$ and unbinds from the filament. The total unbinding
probability per unit time is $\epsilon_0=n_{\rm ad}\epsilon/6$ with 
the number $n_{\rm ad}$ of adjacent non-filament sites which is given 
by $n_{\rm ad}=4$ and $n_{\rm ad}=3$ for filaments in solution and filaments 
immobilized to a surface, respectively. 

At non-filament sites, the motor performs a symmetric random walk and
hops to all neighboring sites with probability $1/6$ per time $\tau$.
This choice of the hopping rate for unbound motor movements implies
that the time scale $\tau$ is given by the diffusion coefficient
$D_\ub$ of unbound motors via $\tau\equiv \ell^2/D_\ub$. If it
reaches a filament site, it binds to the filament with the sticking
probability $\piad$. 

The behavior at the filament end has to be specified separately. We
consider two possibilities: (i) active unbinding of motors at the
filament end where motors at the last filament site make a forward
step with rate $\alpha$ as at the other filament sites, but the latter
step brings them to the forward non-filament neighbor site, so that
their total unbinding probability is $\epsilon_0+\alpha$, and (ii)
thermal unbinding where a motor at the last filament site does not
make a forward step, but has an increased waiting probability
$\gamma'=\gamma+\alpha$. In that case, unbinding occurs with
probability $\epsilon_0$.

The hopping rates can be chosen in such a way that one incorporates
the measured transport properties of single motors such as velocity,
diffusion coefficient, and average walking distance before unbinding
from the filament
\cite{Lipowsky__Nieuwenhuizen2001,Lipowsky_Klumpp2005}.

Finally, motor-motor interactions can easily be incorporated into
these models. In the following we mainly consider the mutual exclusion
of motors from lattice sites. Exclusion is most important at filament
sites (since the motors are strongly attracted to these sites), but in
principle also applies to unbound motors. In the last section of this
article, we consider cooperative binding of motors to the filament. In
that case, the binding and unbinding probabilities depend on the
occupation of the nearest neighbor filament sites.

When considering many-motor systems, one is often interested in the
motor densities and currents profiles rather than the single-motor
trajectories. The quantities of main interest are then the bound and
unbound motor densities $\rho_\bd$ and $\rho_\ub$.  If gradients of
the unbound motor density along the direction parallel to a filament
can be neglected -- either because unbound diffusion is very fast or
because the space available for unbound diffusion is large, so that
motors remain unbound for a long time
before rebinding to the filament -- the unbound density can be treated
as constant. In that case, one obtains a one-dimensional model for the
filament which is coupled to a reservoir of unbound motors with
constant motor density as studied in
\brefs{Parmeggiani__Frey2003,Evans__Santen2003,Klumpp_Lipowsky2004,Klein__Juelicher2005,Nishinari__Chowdhury2005}.
Per unit time $\tau$, a motor on the filament unbinds with probability
$\epsilon_0$ and binding of a motor from the reservoir to an empty
lattice site occurs with probability 
$\pi_0\rho_\ub\equiv n_{\rm ad}\piad\rho_\ub/6$, as shown in
\fig{fig:latticemodel}(b).  This situation will be discussed in
section \ref{jams}.

\section{Motor traffic in tube-like compartments}

The motor traffic through tube-like compartments in which one or
several filaments are aligned parallel to the cylinder axis
represents a simple system which mimics the transport in axons. We
have studied tube-like systems with various kinds of boundary
condition: closed systems
\cite{Lipowsky__Nieuwenhuizen2001,Klumpp__Lipowsky2005,Mueller__Lipowsky2005},
periodic boundary conditions \cite{Klumpp_Lipowsky2003}, open
boundaries coupled to motor reservoirs \cite{Klumpp_Lipowsky2003}, and
half-open systems \cite{Mueller__Lipowsky2005}.

\begin{figure}[tb]
  \centering
  \includegraphics[angle=0,width=0.9\columnwidth]{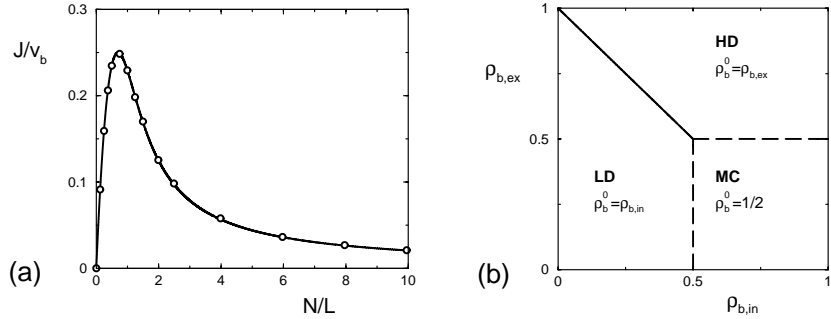}
  \caption{(a) Bound motor current $J$ for a filament in a tube with
    periodic boundary conditions as a function of the number $N/L$ of motor
    particles per length within the tube.  (b) Phase diagram for an open tube
    coupled to motor reservoirs with densities $\rho_{\rm b,in}$ and
    $\rho_{\rm b,ex}$ at the left and right end of the tube. }
  \label{fig:J_Phasendiag}
\end{figure}

The simplest case is given by periodic boundary conditions which can
be solved exactly \cite{Klumpp_Lipowsky2003}. In this case, the
stationary probability distribution is given by a product measure; the
bound and unbound motor densities are constant and satisfy the radial
equilibrium condition
\begin{equation}
  \piad\rho_\ub(1-\rho_\bd)=\epsilon\rho_\bd(1-\rho_\ub)\approx\epsilon\rho_\bd,
\end{equation}
where the last approximation usually holds under experimentally
accessible conditions, where the unbound density is small, but the
bound motor density can be of the order of one motor per binding site.
The bound motor current is given by $J=v_\bd\rho_\bd(1-\rho_\bd)$ with
the bound motor velocity $v_\bd$. As a function of the bound motor
density or of the total number $N$ of motors within the tube, it
exhibits a maximum and decreases for high motor densities due to motor
jamming as shown in \fig{fig:J_Phasendiag}(a).

If the tube is coupled to motor reservoirs at its orifices, the motor
traffic exhibits boundary-induced phase transitions related to those
of the one-dimensional asymmetric simple exclusion process (ASEP)
\cite{Klumpp_Lipowsky2003}. As for the ASEP, a low-density,
high-density and maximal-current phase are present, and correspond to
situations where the bottleneck which limits the transport is given by
the left boundary, the right boundary or the interior of the tube,
respectively. In all three phases, the motor densities are
approximately constant and satisfy radial equilibrium sufficiently far
from the boundaries. The location of the transition lines within the
phase diagram is quite sensitive to the precise choice of the boundary
conditions and can be shifted by tuning the model parameters. A
particularly simple case is obtained if we impose radial equilibrium
at the boundaries. In this case, the phase diagram, which is shown in
\fig{fig:J_Phasendiag}(b), is independent of the motor transport
properties and the geometric parameters of the tube, and the phase
diagram corresponds exactly to the ASEP phase diagram.

\section{Traffic jams on filaments in contact with a large motor reservoir}
\label{jams}

From an experimental point of view, the simplest system, 
for which one can study
molecular motor traffic, is given by one or several immobilized
filaments which are in contact with a solution with a certain motor
concentration. For typical in vitro systems, unbound motor 
diffusion is very fast and the
space available for unbound diffusion is large, so that we can
describe the unbound motors by a constant density $\rho_\ub$. In the
following, we will use dimensionful quantities with units typically
used by the experimentalist, and therefore characterize the unbound
motors by the concentration $c$, which is typically in the nano-molar
range, rather than by the local volume fraction $\rho_\ub$. In these units,
the rate for the binding of an unbound motor to an empty filament site
is given by $\pi_c c$ where $\pi_c$ is the second-order binding rate.
It is related to the binding rate in density units via $\pi_c
c=\pi_0\rho_\ub$ with $\pi_0\equiv n_{\rm ad}\piad/6$ and is most 
conveniently expressed in terms of
the dissociation constant $K_d\equiv\epsilon_0/\pi_c$ which has the
dimension of concentration and is typically of the order of $\sim
100$~nM.

\begin{figure}[tb]
  \centering
  \includegraphics[angle=0,width=\columnwidth]{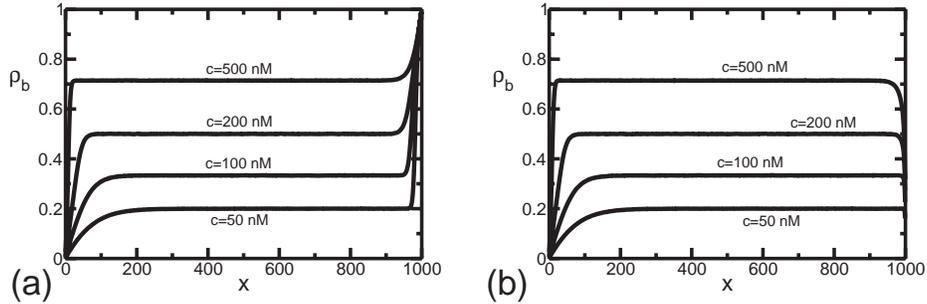}
  \caption{Profiles of the bound motor density $\rho_\bd$ on a filament 
    in contact with a solution with constant unbound motor 
    concentration $c$ as a function of the coordinate $x$ 
    parallel to the filament for (a) thermal and (b) active unbinding at 
    the filament end. Note that a traffic jam only occurs for thermal 
    unbinding at the filament end and that this jam is rather short, of 
    the order of the walking distance. The parameters are as appropriate 
    for kinesin, $\epsilon_0=1/$s, $K_d=200$nM, $v_\bd=1\mu$m/s, and 
    for a microtubule of length $8\mu$m. }
  \label{fig:fil_in_res}
\end{figure}

If the filament is long compared to the motor walking distance, the
bound motor density is constant except for the regions close to the
filament end and given by the equilibrium of the binding/unbinding
dynamics, $\rho_{\rm b}^{(0)}=c/(K_d+c)$ as shown in
\fig{fig:fil_in_res}(a) and (b) for thermal and active unbinding at
the filament end, respectively, using parameters for conventional
kinesin.  Likewise, the current along this part of the filament is
given by $J_0=v_\bd\rho_{\rm b}^{(0)}(1-\rho_{\rm b}^{(0)})$. If the
motors unbind thermally at the filament end, a (rather short) traffic
jam forms at the filament end, where the motors
accumulate. Note that no jam is obtained if the motors unbind actively
as shown in \fig{fig:fil_in_res}(b).

The length of the jam region can be defined as $L_*\equiv L-x_*$ with
the filament length $L$ and $\rho(x_*)=(1+\rho_\bd^{(0)})/2$. An
estimate of $L_*$ can be obtained from the balance of currents
\begin{equation}
  \label{eq:currBal}
  J_0 -J_{\rm end} =\epsilon_0\int_{L-L_*}^L \d x \left[\rho_\bd-\frac{c}{K_d}(1-\rho_\bd)\right],
\end{equation}
where $J_{\rm end}$ is the forward current at the last filament site.
\eq{eq:currBal} leads to the jam length
\begin{equation}
  \label{eq:L*}
  L_*=\uDelta x_\bd \frac{J_0 -J_{\rm end}}{v_\bd}\left[\bar\rho_{\rm b,jam}-\frac{c}{K_d}(1-\bar\rho_{\rm b,jam})\right]^{-1},
\end{equation}
where $\bar\rho_{\rm b,jam}$ is the average bound density in the jam
region and $\uDelta x_\bd$ is the walking distance of the motors as
given by $\uDelta x_\bd\equiv v_\bd/\epsilon_0$.

For thermal unbinding of motors at the filament end, $J_{\rm end}=0$.
If we estimate the density within the jam by the maximal value,
$\bar\rho_{\rm b,jam}\simeq 1$, \eq{eq:L*} leads to
\begin{equation}
 \label{eq:L*_estimate}
  L_*\simeq\uDelta x_\bd \frac{J_0}{v_\bd}\leq \uDelta x_\bd/4
\end{equation}
for the jam length in agreement with simulations which show that
$L_*\simeq\uDelta x_\bd J/v_\bd$ with a prefactor close to one.


This estimate shows that, for filaments in contact with a solution
with constant motor concentration, the jam length $L_*$ is of the
order of the walking distance $\uDelta x_\bd$.  Longer jam lengths can
arise (i) if the unbinding rate $\epsilon$ decreases with
increasing density or (ii) if a gradient in the concentration of
unbound motors is build up
\cite{Lipowsky__Nieuwenhuizen2001,Klumpp__Lipowsky2005} which
increases binding to the filament in the jam region and thus also
increases the last term of \eq{eq:L*}.

\section{Geometry-enhanced traffic jams in closed compartments}

\label{closed}

If filaments are embedded into closed compartments, the motor current
along these filaments leads to the build-up of density gradients
within these compartments
\cite{Lipowsky__Nieuwenhuizen2001,Klumpp__Lipowsky2005,Mueller__Lipowsky2005}.
These gradients are particularly pronounced in tube-like compartments
where all filaments are aligned in parallel and with the same
orientation along the tube axis. For low motor densities, the motors
are essentially localized at that end of the tube towards which their
active movements are directed. If exclusion can be neglected, the
bound and unbound motor densities decrease exponentially if one moves
away from this tube end. The length scale $\xi$ of the exponential
decrease is given by $\xi=D_\ub\phi\epsilon/(v_\bd \piad)$, the ratio
of the walking distance of bound motors and the distance unbound motor
diffuse before rebinding. In this expression, $v_\bd$ and $D_\ub$ are
the bound velocity and the unbound diffusion coefficient of the
motors, respectively, and $\phi$ is the cross-section of the tube. A
constant unbound motor density is a good approximation if $\xi\gg L$,
i.e., for large unbound diffusion coefficients $D_\ub$ and for large
tube radii.

\begin{figure}[tb]
  \centering
  \includegraphics[angle=0,width=0.9\columnwidth]{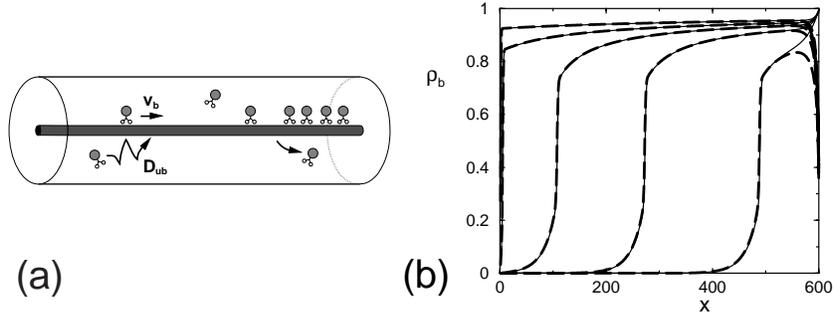}
  \caption{(a) Motor traffic within a closed tube. The current of
    bound motors which move along a filament with velocity $v_\bd$ is
    balanced by a diffusive current of unbound motors which diffuse
    back with the diffusion coefficient $D_\ub$. (b) Corresponding
    profiles of the bound motor density $\rho_\bd$ as a function of
    the coordinate $x$ along the filament. A traffic jam domain at the
    right end of the tube builds up both for thermal and active
    unbinding of motors from the 'last' filament site (solid and
    dashed lines, respectively). With increasing overall motor
    concentration, the crowded domain spreads to the left.}
  \label{fig:endAB}
\end{figure}

If the overall motor density is increased in these systems, the region
in which the motors are localized develops into an extended crowded
domain, see \fig{fig:endAB}. The length $L_*$ of this domain defines
the jam length for these systems. In contrast to the systems discussed
in the previous section, the jam length can be larger than the walking
distance and increases with increasing overall motor concentration
until the crowded domain spreads over the full tube length. In this
crowded domain, the density profiles can approximately be described by
local radial equilibrium. 
For the case of a half-open tube, which is very similar to the closed 
tube, but
more easily accessible to analytical methods, the latter approximation 
shows that the jam length scales essentially as $L_*\sim 1/v_\bd$ 
\cite{Mueller__Lipowsky2005}
rather than $L_*\sim \uDelta x_\bd\sim
v_\bd$ as for a filament in contact with a constant unbound motor
density. The jam length is given by
\begin{equation}
  \label{eq:halfopentube}
  L_*=\frac{\phi D_\ub\piad}{v_\bd\epsilon}G(\epsilon/\piad,\rho_{\rm b,in}),
\end{equation}
where $G$ is a function of the ratio of the binding and unbinding
probabilities and of the bound motor density $\rho_{\rm b,in}$ in the
reservoir to which the tube is coupled at its open end
\cite{Mueller__Lipowsky2005}. If the boundary density is sufficiently 
close to
one, $G$ behaves as $G\approx -\ln(1-\rho_{\rm b,in})$ and the jam 
length diverges logarithmically with $1-\rho_{\rm b,in}$. For the closed 
tube, $G$ is determined by an integral constraint which fixes the total 
number of motors within the tube.


In addition, the traffic jam is present for both thermal and active
unbinding at the filament end \cite{Klumpp__Lipowsky2005} as shown in
\fig{fig:endAB}(b). This means that the crowded domains are due to a
combination of the motor behavior at the last filament site and the
motor accumulation in the region close to the filament end. The latter
accumulation is strongly geometry-dependent.

We have also studied centered or aster-like filament systems
\cite{Klumpp__Lipowsky2005}. In this case, the accumulation of motors
in the center of an aster is much weaker than in tube-like systems
and, in fact, determined by a power law rather than by an exponential.
As for filaments in contact with a reservoir with constant unbound
motor density, traffic jams are obtained only for thermal unbinding at
the filament end, but not for active unbinding. In addition, when the
overall motor density is increased, the traffic jams remain short in
this case. The main effect of an increase in the overall motor density
is a flattening of the density profile.

\section{Symmetry breaking and traffic lanes in systems with two motor
  species}

Each molecular motor moves either towards the plus- or towards the
minus-end of the corresponding filament, but different types of motors
move into opposite directions along the same filament. In this
situation, cooperative binding of the motors to the filament -- in
such a fashion that a motor is more likely to bind and less likely to
unbind next to a bound motor moving in the same direction, while it is
less likely to bind and more likely to unbind next to a motor with
opposite directionality -- leads to spontaneous symmetry breaking
\cite{Klumpp_Lipowsky2004}: If the motor--motor interactions, which we
characterize by a single interaction parameter $q$, are stronger than
a certain critical interaction strength $q_c$, one motor species
occupies the filament, while the other one is largely excluded from
it. This symmetry breaking has been found both for tube-like
compartments with periodic boundary conditions and for systems with a
constant unbound motor density. In the latter case, symmetry breaking
occurs, independent of the choice of the boundary conditions provided
that the system size or the filament length is large compared to the
motors' walking distance. Note that, in contrast to the previously
reported example for symmetry breaking in a driven diffusive system,
the 'bridge model' \cite{Evans__Mukamel1995}, the symmetry breaking
here is not boundary-induced.

Symmetry breaking has two interesting consequences. First, it implies
that for $q>q_c$ there is a discontinuous phase transition if the
relative concentrations of the two motor species are varied. This
transition is accompanied by hysteresis, which is again not
boundary-induced, in contrast to the hysteresis which was reported
recently for another driven diffusive system
\cite{Rakos__Schuetz2003}. Second, if several filaments are aligned in
parallel and with the same orientation, this symmetry breaking leads
to the spontaneous formation of traffic lanes for motor traffic with
opposite directionality \cite{Klumpp_Lipowsky2004}.

\section{Concluding remarks}

The traffic phenomena in systems with many molecular motors can be
described by stochastic lattice gas models which are similar to
asymmetric exclusion processes, but have the additional property that
the motors can unbind from the filamentous track and diffuse in the
surrounding fluid. These systems exhibit a variety of cooperative
phenomena and, in addition to their importance for our understanding
of the traffic within cells and for prospective applications in
nanotechnology, provide promising model systems for the experimental
study of driven diffusive systems.\\



\appendix

\end{document}